\documentclass[conference]{IEEEtran}
\IEEEoverridecommandlockouts
\usepackage{subfigure}
\usepackage{amsmath,amssymb,amsfonts}
\usepackage{algorithmic}
\usepackage{algorithm}
\usepackage{array}
\usepackage{textcomp}
\usepackage{stfloats}
\usepackage{url}
\usepackage{verbatim}
\usepackage{graphicx}
\usepackage{cite}
\usepackage{hyperref}
\usepackage{booktabs}
\usepackage{wrapfig}
\hypersetup{hypertex=true,
            colorlinks=true,
            linkcolor=blue,
            anchorcolor=blue,
            citecolor=blue}
\DeclareRobustCommand*{\IEEEauthorrefmark}[1]{%
    \raisebox{0pt}[0pt][0pt]{\textsuperscript{\footnotesize\ensuremath{#1}}}}

\def\BibTeX{{\rm B\kern-.05em{\sc i\kern-.025em b}\kern-.08em
    T\kern-.1667em\lower.7ex\hbox{E}\kern-.125emX}}

\begin{document}

\title{Design and Prototyping of Transmissive RIS-Aided Wireless Communication }

\author{
\IEEEauthorblockN{ Jianan zhang\IEEEauthorrefmark{1}, Rujing Xiong\IEEEauthorrefmark{1}, Junshuo Liu\IEEEauthorrefmark{1}, Tiebin Mi\IEEEauthorrefmark{1}, Robert Caiming Qiu\IEEEauthorrefmark{1}} 
\IEEEauthorblockA{\IEEEauthorrefmark{1}School of Electronic Information and Communications, Huazhong University of Science and Technology, Wuhan 430074, China}
Emails: \{zhangjn, rujing, junshuo\_liu, mitiebin, caiming\}@hust.edu.cn
}
\maketitle

\begin{abstract}
Reconfigurable Intelligent Surfaces (RISs) exhibit promising enhancements in coverage and data rates for wireless communication systems, particularly in the context of 5G and beyond. This paper introduces a novel approach by focusing on the design and prototyping of a transmissive RIS, contrasting with existing research predominantly centered on reflective RIS. The achievement of 1-bit transmissive RIS through the antisymmetry configuration of the two PIN diodes, nearly uniform transmission magnitudes but inversed phase states in a wide band can be obtained. A transmissive RIS prototype consisting of 16 $\times$ 16 elements is meticulously designed, fabricated, and subjected to measurement to validate the proposed design. The results demonstrate that the proposed RIS unit cell achieves effective 1-bit phase tuning with minimal insertion loss and a transmission bandwidth of 3 dB exceeding $20\%$ at 5.8GHz. By dynamically modulating the quantized code distributions on the RIS, it becomes possible to construct scanning beams. The experimental outcomes of the RIS-assisted communication system validate that, in comparison to scenarios without RIS, the signal receiving power experiences an increase of approximately 7dB when RIS is deployed to overcome obstacles. This underscores the potential applicability of mobile RIS in practical communication.
\end{abstract}


\section{Introduction}
\IEEEPARstart{R}{econfigurable} Intelligent Surfaces (RISs) have attracted significant interest in the recent years from both academia and industry~\cite{liaskos2018new}~\cite{basar2019wireless}. Recently, RIS has attracted great attention for its capability of making full use of spatial resources and
improving spectral efficiency by smartly manipulating electromagnetic (EM) environments
~\cite{tang2020wireless}. Practically speaking, an RIS can be built using artificial electromagnetic metamaterial, which consists of periodic arrangements of specifically designed subwavelength-sized structural elements~\cite{2009Metamaterials}. Such metamaterials have unique electromagnetic properties that do not exist in nature~\cite{cui2017information}, such as, negative refraction~\cite{shelby2001experimental}, perfect absorption, and anomalous reflection/scattering. The intelligent reconfiguration of the phase states of RIS elements enables the formation of arbitrary beam shapes at a relatively low cost and power consumption. RIS technology offers substantial advantages in extending network coverage, enhancing capacity, conserving power, and mitigating interference in beyond 5G (B5G) networks. Consequently, numerous signal processing techniques and experimental studies have been conducted to showcase the efficacy of RIS technology~\cite{dai2020reconfigurable}
~\cite{huang2019reconfigurable}.

However, current research efforts predominantly focus on the application of reflective Reconfigurable Intelligent Surfaces (RIS)~\cite{dai2020reconfigurable}~\cite{huang2019reconfigurable}~\cite{pei2021ris}~\cite{trichopoulos2022design}. This emphasis on reflective RIS poses the risk of creating undesirable coverage gaps within cellular networks, ultimately limiting the overall applicability of RIS technology. Notably, reflective RIS introduces a constraint where both the base station and user must be situated on the same side of the RIS, imposing additional geographical restrictions on the network's physical topology~\cite{wang2023doppler}. For instance, facilitating communication between a transmitter outside a vehicle and a receiver inside it becomes challenging with reflective RIS. To address these limitations and broaden the scope of RIS applications, increasing attention is being directed toward the concept of transmissive RIS~\cite{docomodocomo}. In contrast to reflective RIS, where signals are reflected, transmissive RIS enables signals to pass through, forming directional beams. This unique characteristic of transmissive RIS holds the potential to effectively fill the coverage gaps left by reflective RIS, overcoming challenges associated with constrained communication scenarios. The rising interest in transmissive RIS underscores its significance in mitigating limitations associated with reflective RIS and expanding the application possibilities within wireless communication networks.

The phase reconfigurability is the most essential capability for a transmissive RIS. Solid-state electronic devices are commonly integrated in each constituent RIS element to dynamically control its phase response. Continuous phase shifts can be realized using analog-type devices like varactor diodes, PIN diodes, Micro-Electro-Mechanical System, ane etc (MEMS) switches~\cite{sazegar2012beam}~\cite{huang2015using}~\cite{clemente20121}~\cite{ di2016circularly}~\cite{ nguyen2018unit}.

In this paper, an efficient transmissive metasurface is presented to generate dynamically scanning beams. The designed metasurface unit can provide very high transmission efficiency, and the 1-bit phase tuning of the unit is obtained by integrating two PIN diodes in the unit structure for current inversion. Through the modulation of the switchable status on PIN diodes by using programmable bias circuit, dynamical scanning beams can be created. The simulation and experimental results verify the effectiveness of the proposed design. Finally, the transmitted RIS is integrated into the wireless communication system, which shows good capability in the scenario of through-the-wall. 
The main contributions of our work are as follows:

\begin{itemize}
\item  The achievement of 1-bit transmissive RIS through the antisymmetry configuration of the two PIN diodes, nearly uniform transmission magnitudes but inversed phase states in a wide band can be obtained.
\item Furthermore, a prototype with 16 $\times$ 16 RIS elements and a logic circuit control board is fabricated and measured. The measured results show that the prototype can achieve $180^{\circ}$ phase shift. Moreover, the measured radiation performances for scan angles are presented as well, demonstrating the beamsteering capability.
\item Finally, a transmissive RIS-aided wireless communication prototype is developed to demonstrate the performance in the scenario of through-the-wall. The
experimental results verify that transmitted RIS enhanced the gain by about 7dB.
\end{itemize}
The rest of this paper is organized as follows. Section~\ref{section2} describes the structure and simulated performance of the element of the designed transmissive RIS. Section~\ref{section3} provides the experimental results for the 16 $\times$ 16 elements RIS prototype.
\section{RIS design }\label{section2}
\subsection{ unit cell design }
\begin{figure}[!htbp]
  \centering
  \subfigure[]{
    \label{F:Fig:schematic}
    \includegraphics[width=1\columnwidth]{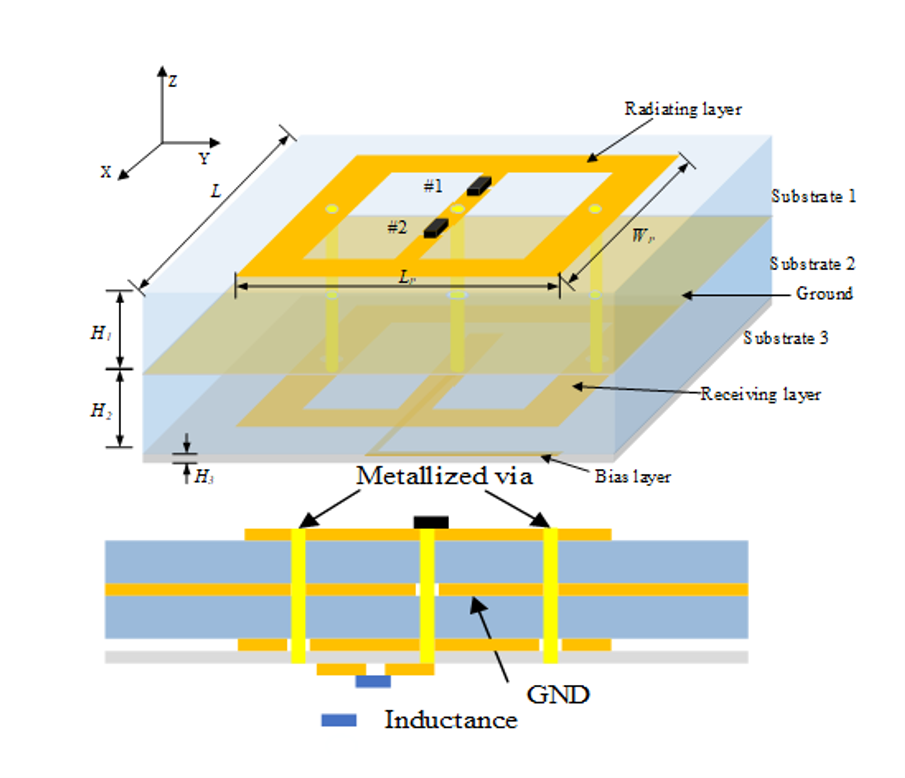}}
  \subfigure[]{
    \label{F:Fig:active layer}
    \includegraphics[width=0.47\columnwidth]{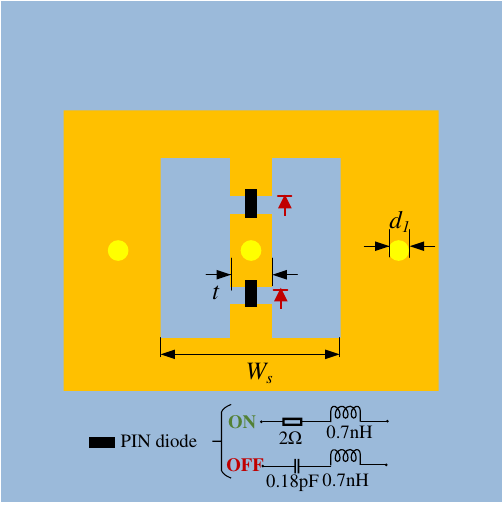}}
\subfigure[]{
    \label{F:Fig:passive layer}
    \includegraphics[width=0.47\columnwidth]{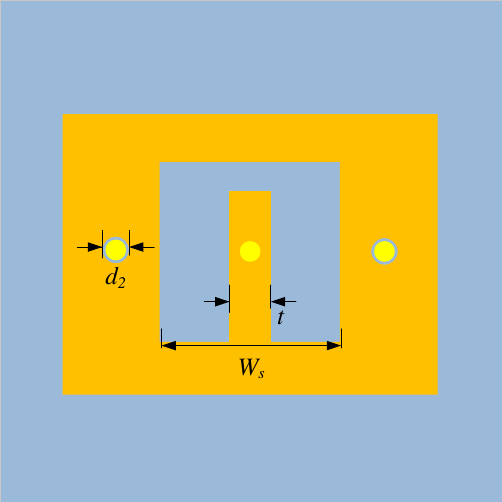}}
  \caption{Geometry of the designed metasurface element, (a) schematic and side view, (b) Radiating layer and (c) Receiving layer}
\label{F:Fig:cell structure}
\end{figure}

The geometry of the RIS unit-cell is shown in Fig.~\ref{F:Fig:cell structure}, which is made up of four copper layers supported by three substrate layers. The copper layers encompass, from top to down, the radiating layer, the ground plane, the receiving layer, and the bias layer. The radiating layer is composed of a rectangular annular patch and a rectangular patch with two PIN diodes (\#1 and \# 2) integrated at the junctions along the $x$-axis. The outer  rectangular annular patch is connected with the ground plane by two metallized via-holes symmetrically on the horizontal center line through the receiving layer, while the rectangular inner patch is connected with the receiving layer by a metallized via-hole at the center through the ground, which consists of the same size rectangular annular patch loaded by a U-slot. Besides, the receiving layer is also connected with the bias layer through center metallized via-holes for biasing purposes, and the bias layer is integrated with a 15 nH inductor to isolate the high frequency signals from the patches to the bias lines. The substrate 1 and substrate 2 layers are designed using a F4B$(\varepsilon_r = 3.5, \tan \delta = 0.001, H_1,H_2= 2 mm)$, while the bias layer is with a dielectric substrate 3 $(\varepsilon_r = 4.3, \tan \delta = 0.025, H_1,H_2= 0.2 mm)$. The detailed dimensions of the proposed unit are also provided in table~\ref{tab:table1}.
\begin{table}[!t]
\caption{Parameters for the transmissive unit cell in millimeters\label{tab:table1}}
\centering
\begin{tabular}{|c|c|c|c|c|c|c|c|c|c|c|c|c|}
\hline
Parameter & $L$ & $L_p$ & $W_p$ & $H_1$ & $H_2$ & $W_s$ & $t$ & $d_1$ & $d_2$\\
\hline
Value & 18 & 15 & 11 & 2 & 2 & 8 & 2 & 0.3 & 0.5\\
\hline
\end{tabular}
\end{table}

\begin{figure}[!htbp]
  \centering
  \subfigure[]{
    \label{F:Fig:current_0_state}
    \includegraphics[width=0.45\columnwidth]{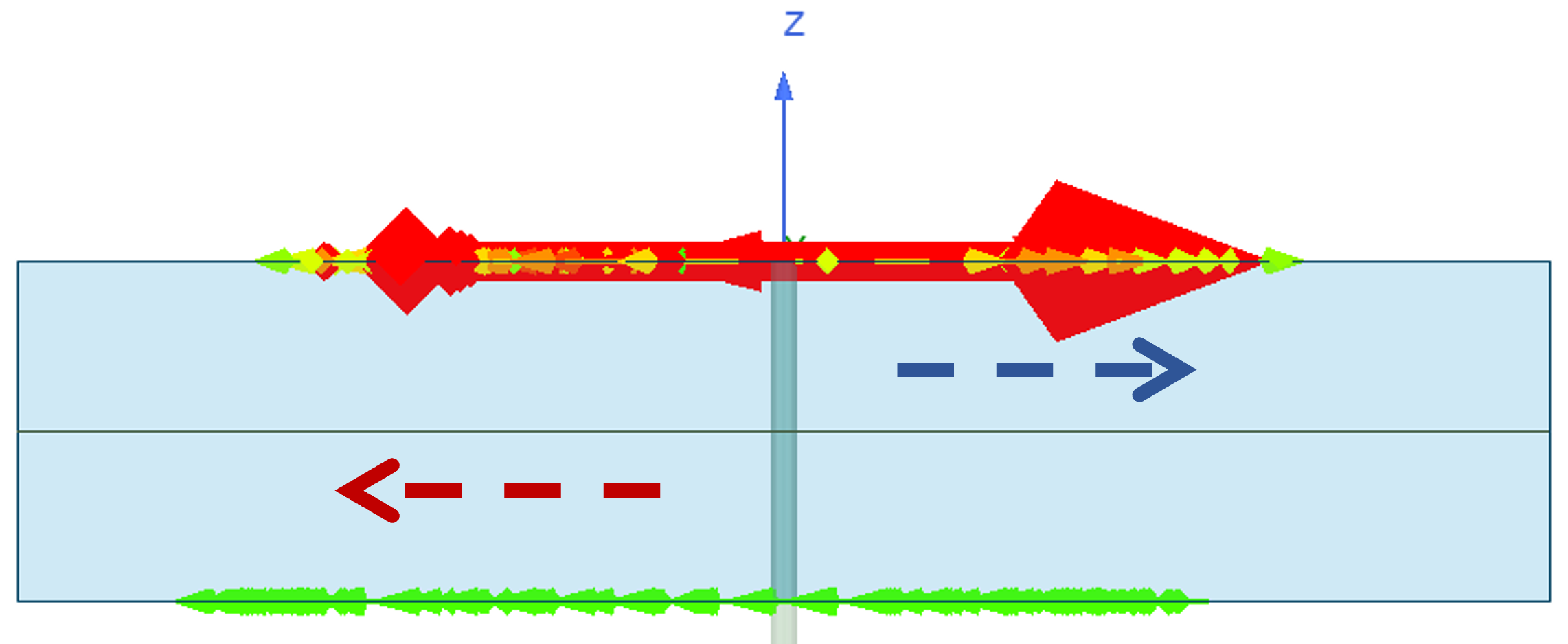}}
  \subfigure[]{
    \label{F:Fig:current 1_state}
    \includegraphics[width=0.49\columnwidth]{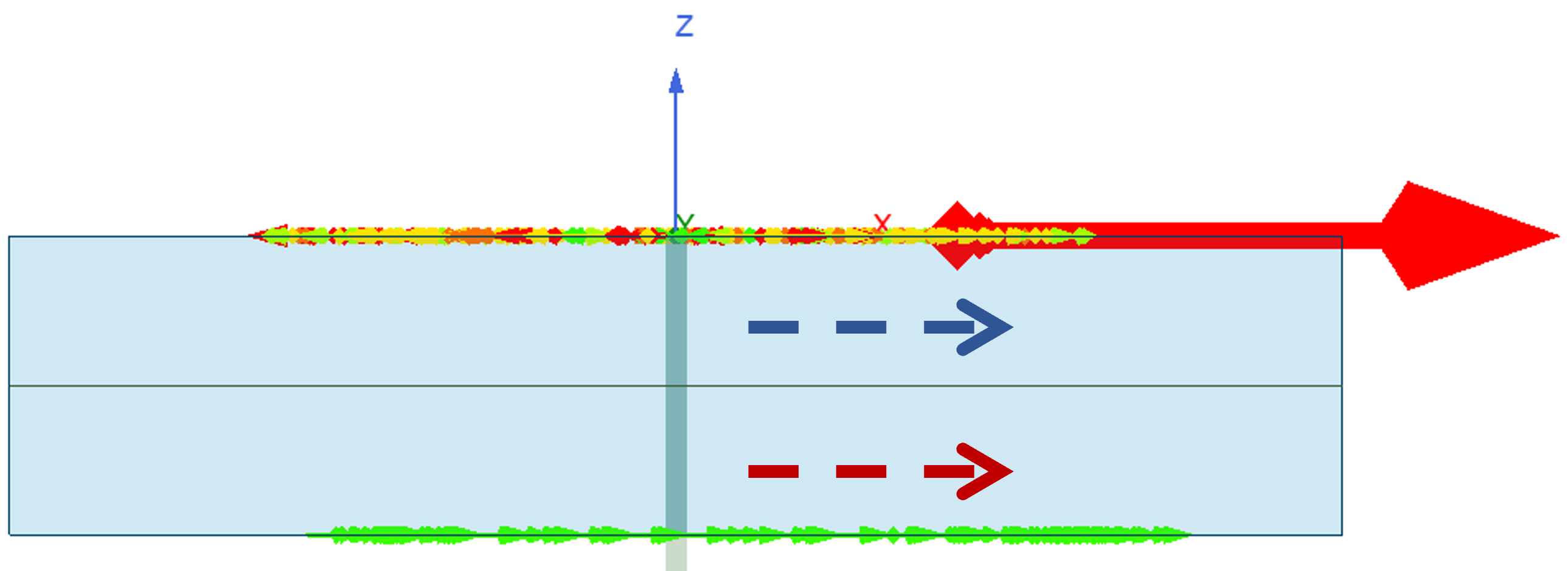}}
  \caption{Current distribution on the unit cell, (a) 0-state, (b) 1-state}
\label{F:Fig:Current distribution}
\end{figure}

The two PIN diodes integrated on the unit are placed with antisymmetry configuration as shown in Fig~\ref{F:Fig:active layer}, resulting in the opposite PIN biased states for any of the two operation states. When PIN diode \#1 is switched off and PIN diode \#2 switched on, the proposed reconfigurable unit is operating at the 0-state. Correspondingly, the 1-state can be acquired with PIN diode \#1 switched on and PIN diode \#2 switched off. Fig.~\ref{F:Fig:Current distribution} depicts the induced current distribution on the radiating layer and receiving layer in
two states. Since the current directions of the receiving layer in  0-state and  1-state are opposite, a phase shift of transmitted wave on the unit cell can be obtained between two cases. That is, the two coaxial patches on the radiating layer would be concatenated at the upper or nether junction for the two coding states, respectively, leading to a current inversion for the two operation coding states and obtaining the inverting transmission coefficients with nearly uniform magnitudes but inversed phase states in a wide frequency band. The PIN diode SMP1345-079LF is adopted to acquire lower unit insertion losses within the designed frequency band, which is modeled by the lumped components in series for ON/OFF states, respectively. For the ON state with a forward
dc-biasing voltage, a series of lumped resistance (R = 2 $\Omega$) and inductance (L = 0.7 pH) is adopted for the PIN diode; while for the OFF state with a reverse dc-biasing voltage, a capacitance (C = 0.18 pF) and inductance (L = 0.7 pH) is employed. The specific equivalent circuits are shown in Fig.~\ref{F:Fig:active layer}. 

\begin{figure}[!htbp]
  \centering
\includegraphics[width=1\columnwidth]{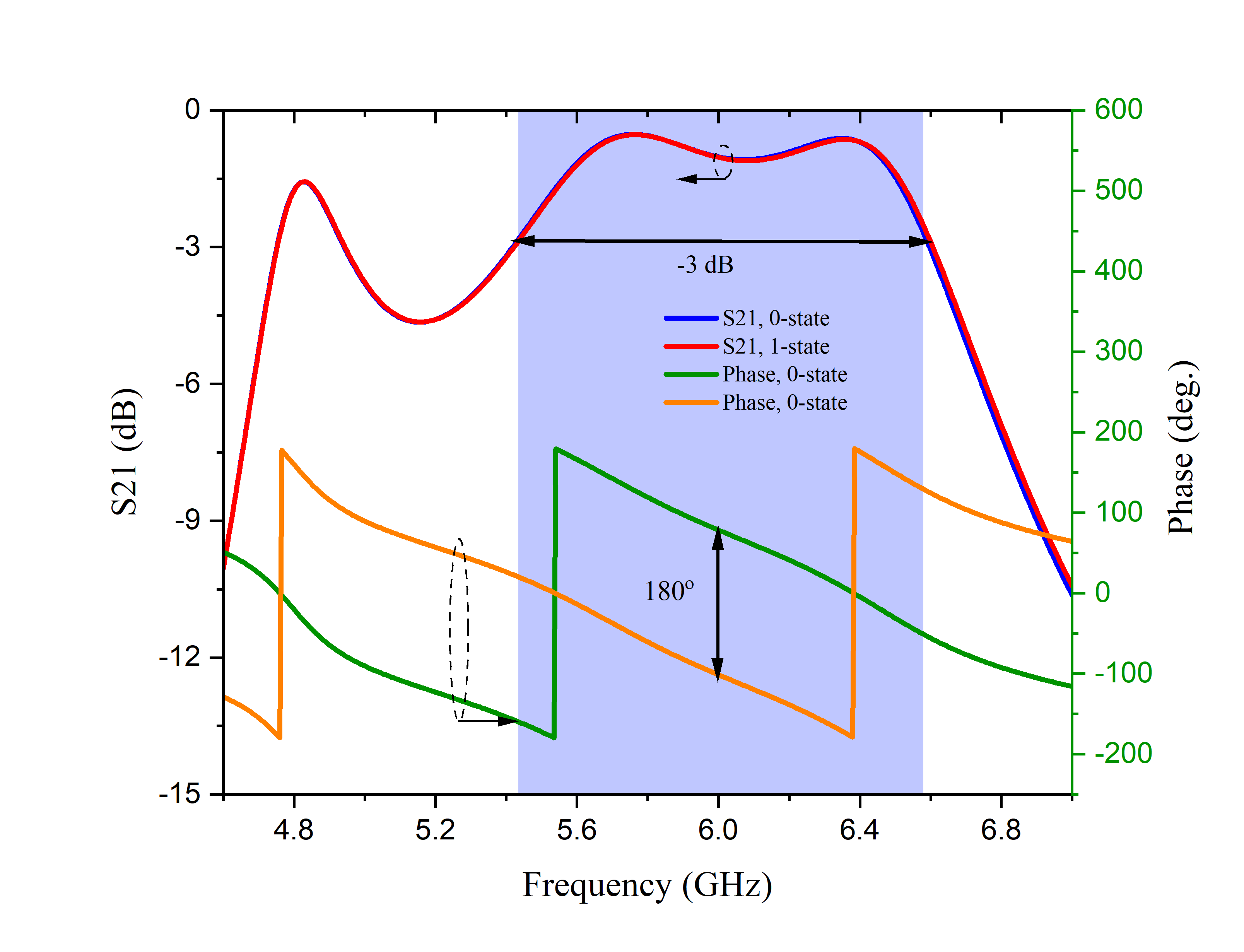}
\caption{Simulated results of unit-cell in different states}
\label{F:Fig:cell S21}
\end{figure}

The electromagnetic simulation of the proposed transmissive unit is carried out with help of the commercial software package CST Microwave Studio by using unit cell boundary conditions along with Floquet-port excitations. The simulated scattering coefficients with both magnitude and phase for the two phase states of the unit cell are shown in Fig.~\ref{F:Fig:cell S21}. When the proposed unit is operating at the 0-state and 1-state, the $S_{21}$ at 5.4 GHz to 6.6 GHz is greater than -3 dB. The simulated 3 dB transmission bandwidths are over 20\%, with the minimum insertion losses of only 0.5 dB at 5.8 GHz. It is worth mentioning that the phase curves of the two codeing cases are basically parallel throughout the frequency band, and the phase difference is about $180^{\circ}$.

\subsection{ Array design and fabrication} 
\begin{figure}[!htbp] 
  \centering
  \subfigure[]{
    \label{F:Fig:RIS PCB}
    \includegraphics[width=0.44\columnwidth]{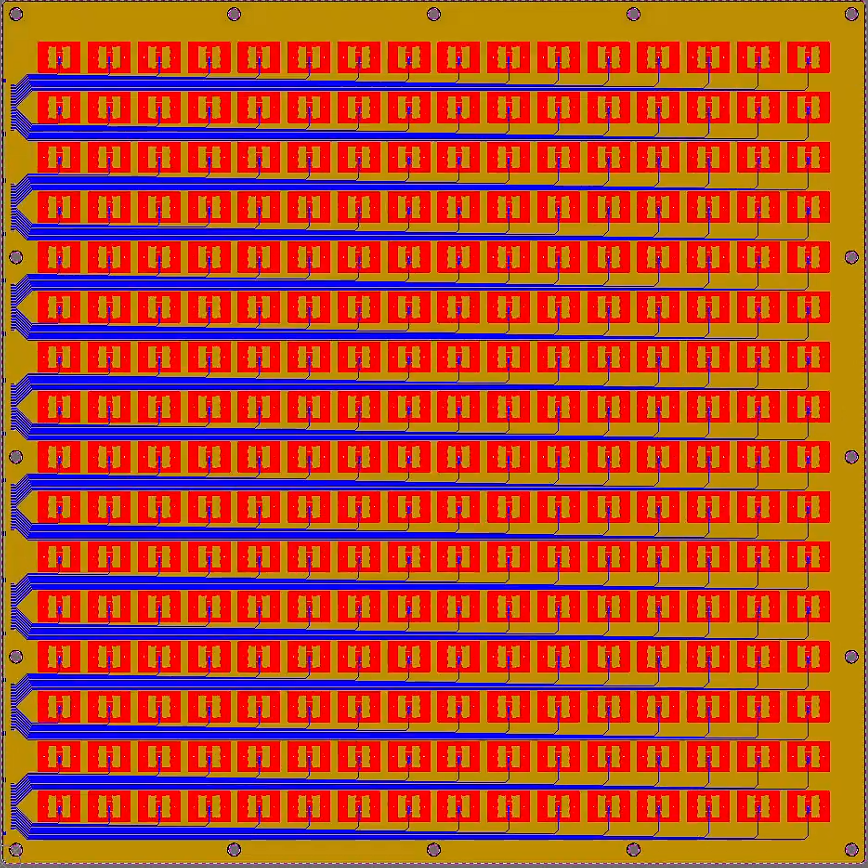}}
  \subfigure[]{
    \label{F:Fig:control board}
    \includegraphics[width=0.52\columnwidth]{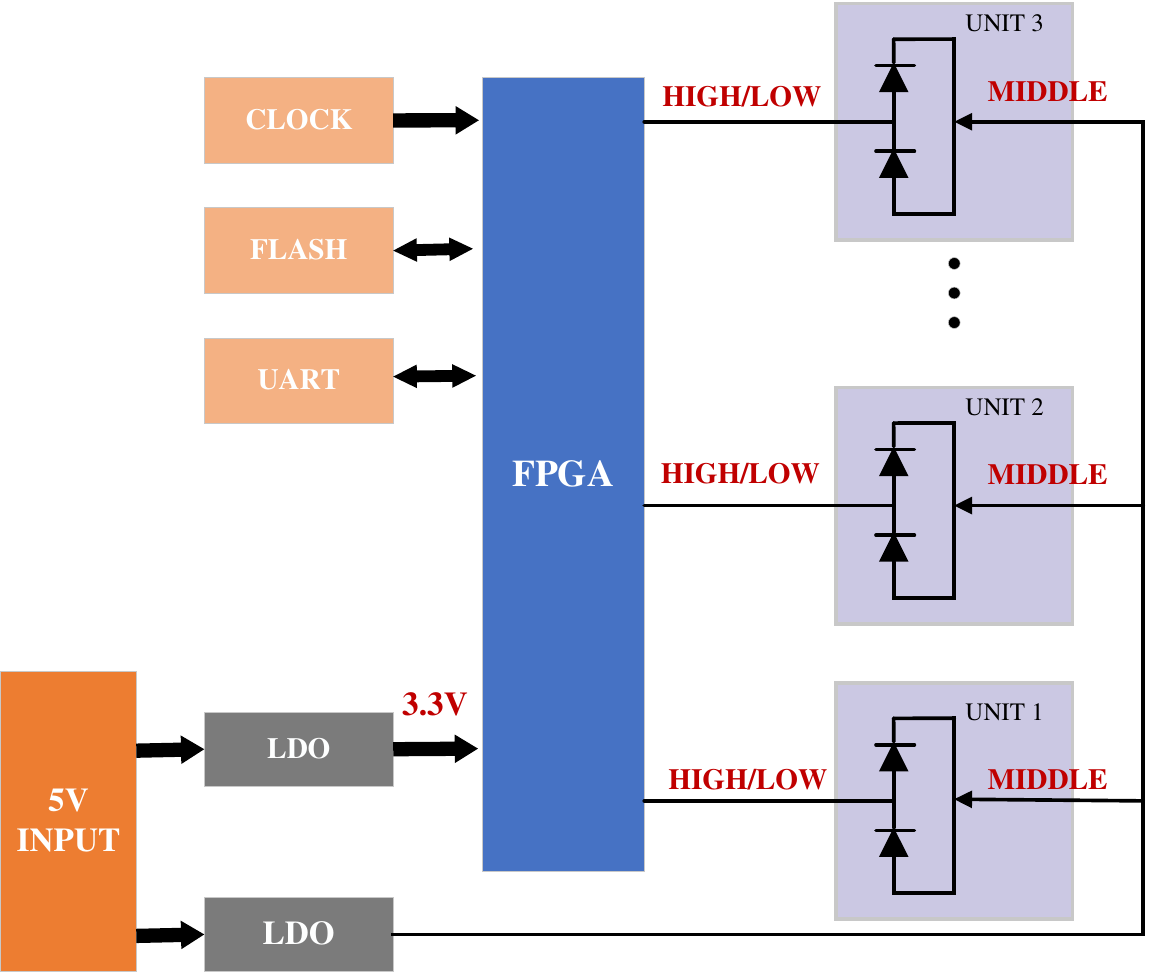}}
 \subfigure[]{
    \label{F:Fig:fabricated}
    \includegraphics[width=0.52\columnwidth]{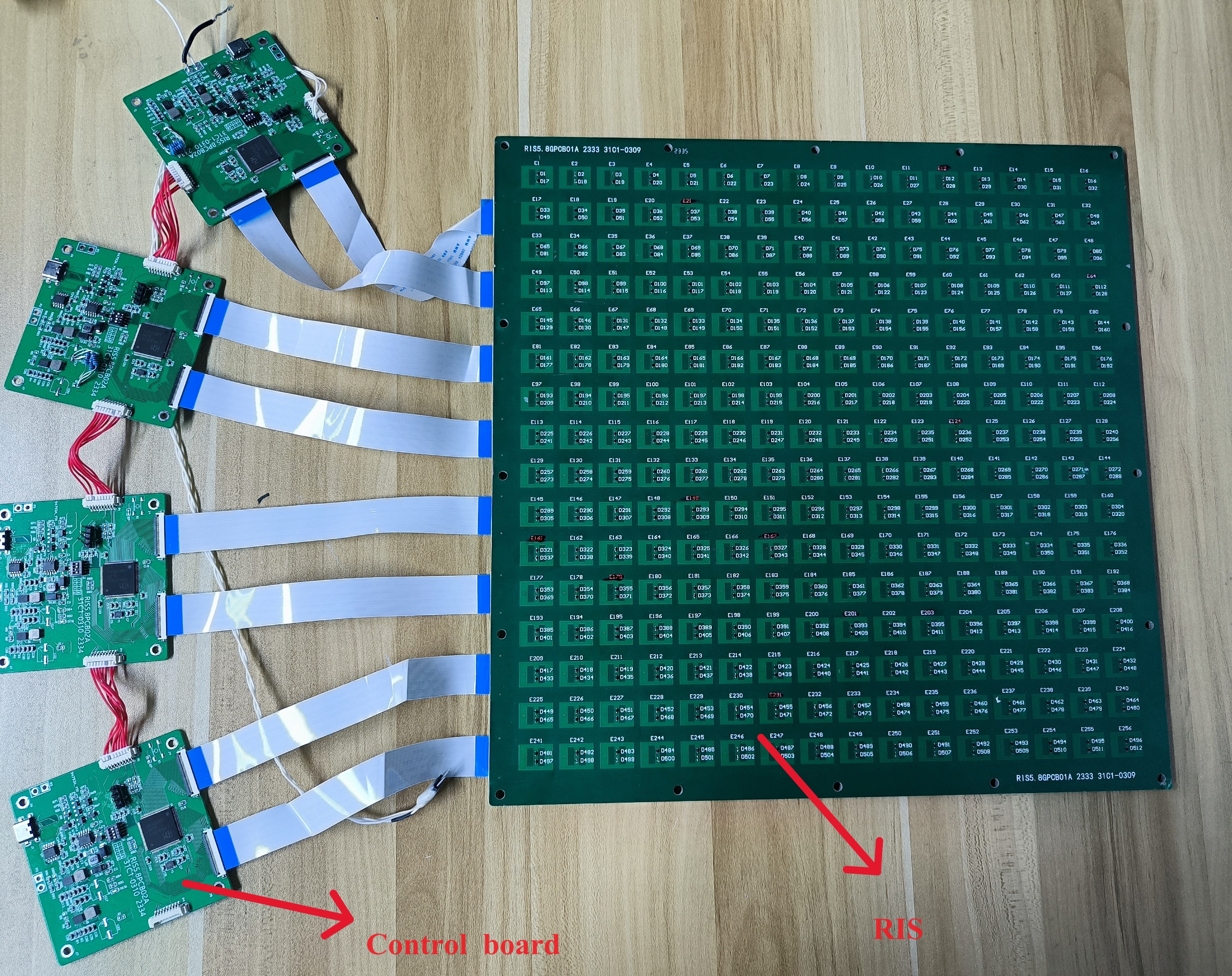}}
  \caption{a) The bias layer network layout. b) Schematic diagram of the steering-logic board circuit design. (c) the fabricated transmissive RIS prototype.}
\label{F:Fig: board circuit}
\end{figure}

In this section, a transmissive RIS prototype composed of 16 × 16 proposed unit cells is designed, fabricated, and measured to experimentally validate the performance of the proposed RIS unit cell, where the bias lines should be arranged on the upper and lower edges of the unit cell and minimize the bias-layer interference, as shown in ~\ref{F:Fig:RIS PCB}.

The transmissive RIS encompasses a total of 512 PIN diodes and the state of each unit cell is modulated through the bias layer network, which are interfaced to eight $1\times 34$ pins connectors. The steering-logic board is also designed and fabricated, and the schematic diagram of the circuit design is shown in Fig.~\ref{F:Fig: board circuit}. The whole control system is powered by a voltage input supply (5 V), and an extra low dropout regulator (LDO) provides an external reference voltage for the metasurface ground plane (marked as MIDDLE).
with an overall dimension of 300 × 300 $mm^2$.

The photographs of the fabricated RIS prototype are presented in Fig.~\ref{F:Fig:fabricated}. Taking advantage of the PCB technology, the RIS is easy to fabricate. To achieve the goal of phase reconfigurability, each element of the array needs to be controlled independently. There are 256 biasing lines since
each element has one biasing lines to apply DC voltages to two PIN diodes. Four logic circuit controlling boards are attached to supply independent DC voltages.
\section{Experimental results} \label{section3}
In this section, we present the results from several experiments that were made using our prototype to verify the performance of this proof-of-concept. First, we test the electromagnetic  transmission coefficients of the RIS. Next We test the radiation pattern in an anechoic chamber to measure the
beamforming function of RIS. Finally, RIS is embedded into the wireless communication system to test its through-wall ability.
\subsection{Transmission Coefficients}
\begin{figure}[!htbp]
  \centering
\includegraphics[width=0.9\columnwidth]{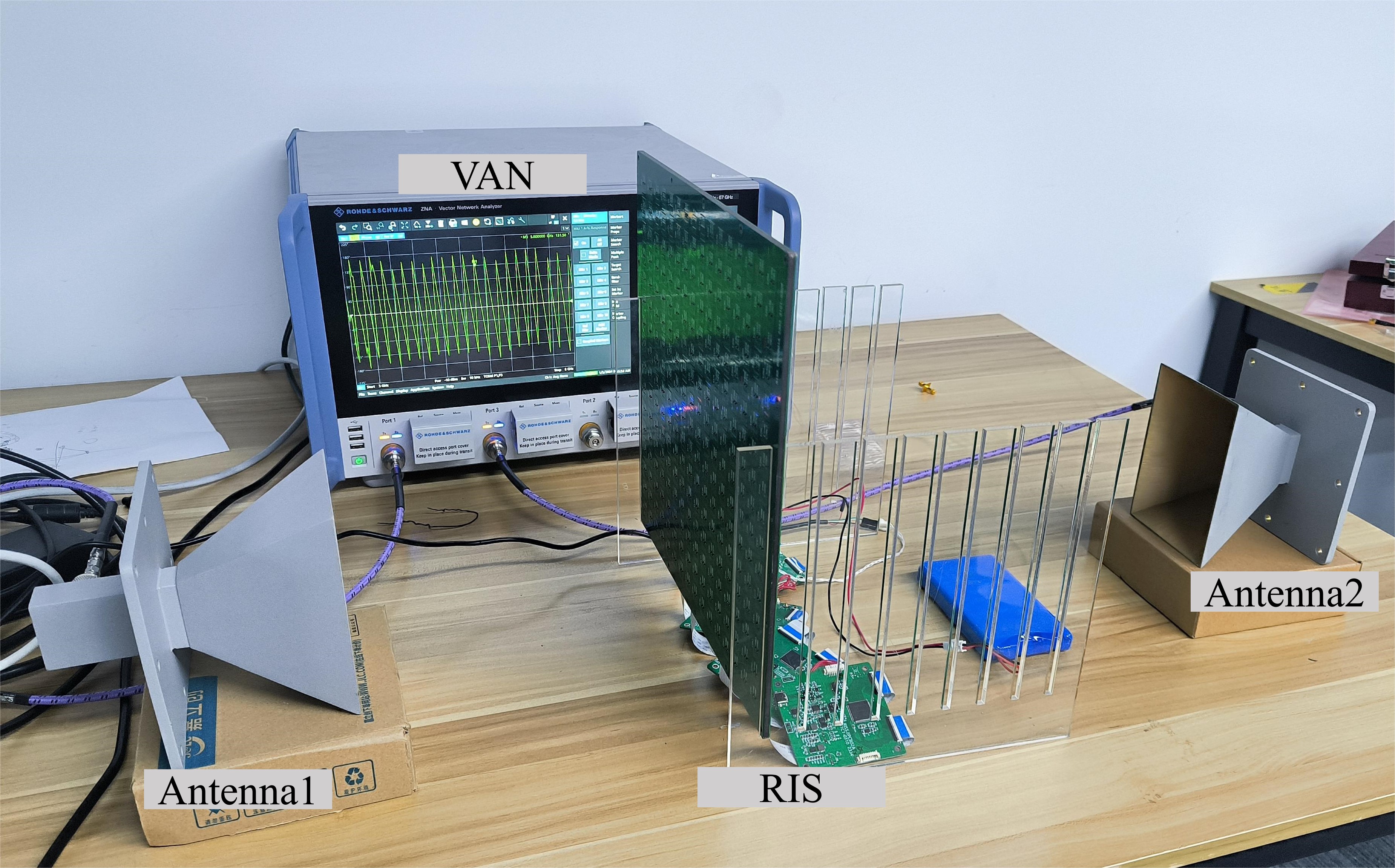}
\caption{S21 test environment}
\label{F:Fig:S21_setting}
\end{figure}

\begin{figure}[!htbp]
  \centering
\includegraphics[width=1\columnwidth]{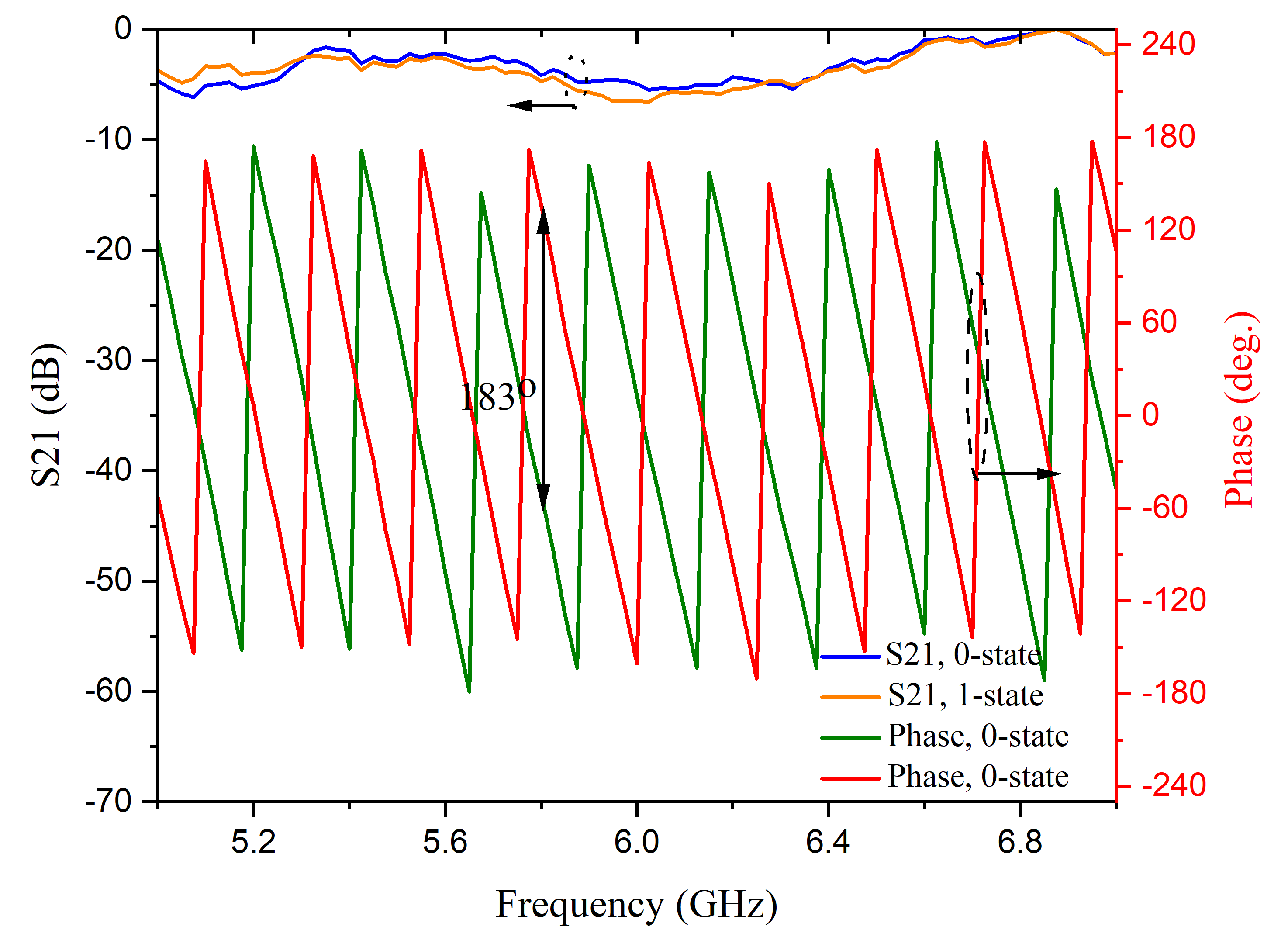}
\caption{ Measured S21 of RIS for both states }
\label{F:Fig:S21_Mea}
\end{figure}

To evaluate the accuracy of the array design simulations, we measure the transmission coefficient $(S_{21})$ of RIS for both states using a vector network analyzer (VNA). As shown in Fig.~\ref{F:Fig:S21_setting}, two antennas are placed on both sides of RIS 0.5 m away from the RIS and a sinusoidal signal at 5.8 GHz is used for the excitation. As plotted in Fig.~\ref{F:Fig:S21_Mea}, the experimental results show that the amplitude is -3 dB and the phase difference is around $183^{\circ}$ for a wide frequency range around 5.8 GHz which is in accordance to the simulations. The amplitude is also normalized and the maximum value is 0 dB.
\subsection{Beamforming Performance}
\begin{figure}[!htbp]
  \centering
\includegraphics[width=0.8\columnwidth]{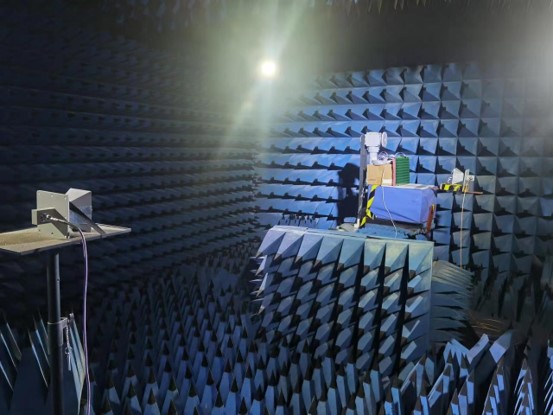}
\caption{The measurement setup in the microwave anechoic chamber.}
\label{F:Fig:microwave anechoic chamber}
\end{figure}

\begin{figure}[!htbp]
  \centering
\includegraphics[width=1\columnwidth]{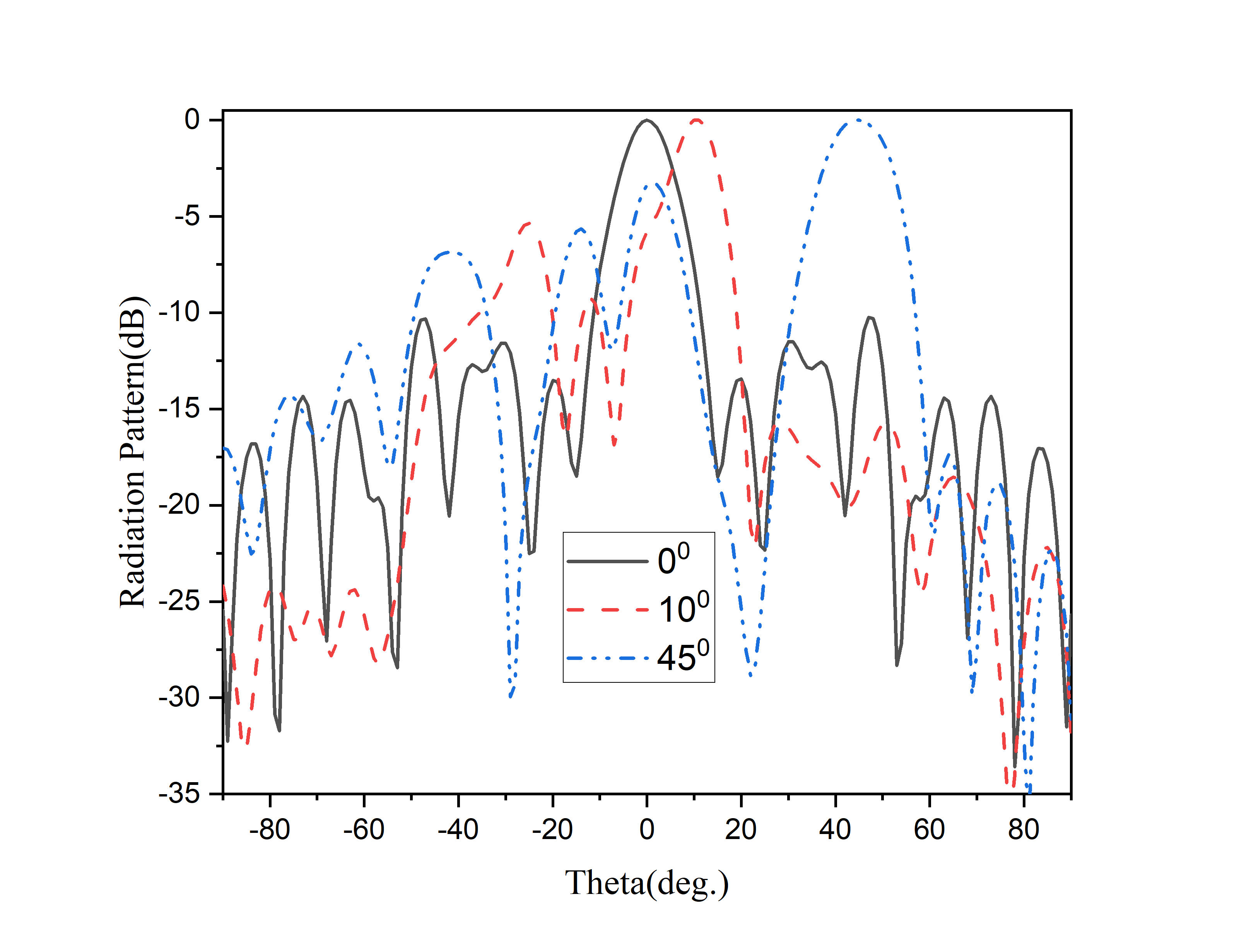}
\caption{Measured radiation patterns of scanned beams at 5.8 GHz}
\label{F:Fig:beamforming}
\end{figure}

Next, we test the beamforming performance of the fabricated RIS. The experiment is carried out in a microwave anechoic chamber as shown in Fig.~\ref{F:Fig:microwave anechoic chamber}. The RIS is space-fed by a linearly polarized horn with an optimized distance of 260 mm.  
To stimulate the scanning beams codebooks through the transmissive metasurface, the theoretical phase compensation $\phi$  of each unit should satisfy the following equation~\cite{nayeri2018reflectarray}.
\begin{equation}
\label{deqn_ex1a}
\phi(x_i, y_i) = \frac{2 \pi}{\lambda}(d_i - \sin \theta_0(x_i \cos \varphi_0 + y_i \sin \varphi_0)).
\end{equation}
where $\lambda$ is the free-space wavelength at 5.8 GHz, $(x_i, y_i)$ is the position of each element, $(\theta_0, \varphi_0)$ is the main beam direction, and $d_i$ is the distance from the feed phase center to the $i^{th}$ element.
Three scanning beams along the axis direction are created as validations through the transmissive programmable metasurface. The result is shown in Fig.~\ref{F:Fig:beamforming}, the measured scanned beams in the angular $0^{\circ}$, $10^{\circ}$ and $45^{\circ}$ show accurate beam pointing and well-defined patterns.
\subsection{Through-wall experiment}
In this section, a transmissive RIS-aided wireless communication prototype is set up to further measure the performance of the designed transmissive RIS in practical communication scenarios.

To assess the efficacy of RIS in overcoming obstacles, we conduct experiments within an office setting, depicted in Figure~\ref{F:Fig:wall}. The prototype comprises the transmitter side, the RIS, and the receiver side. Both the transmitter and the receiver employ the software-defined radio (SDR) platform. The transmitter is situated within the office, while the receiver is positioned in the corridor with a 30 mm concrete barrier between them, and the RIS positioned against the wall. Let $d_1$ denote the distance between the Tx antenna and the RIS, and Table ~\ref{tab:table2} shows the receiving power at different $d_1$. When $d_1$ is 0.5m, 0.8 m and 1.65m respectively, the gain of with RIS is 8dBm, 7dBm and 6dBm higher than that of without RIS. The experimental results show that RIS has a good performance in penetrating obstacles.

\begin{figure}[!htbp]
  \centering
\subfigure[]{
    \label{F:Fig:wall}
    \includegraphics[width=0.52\columnwidth]{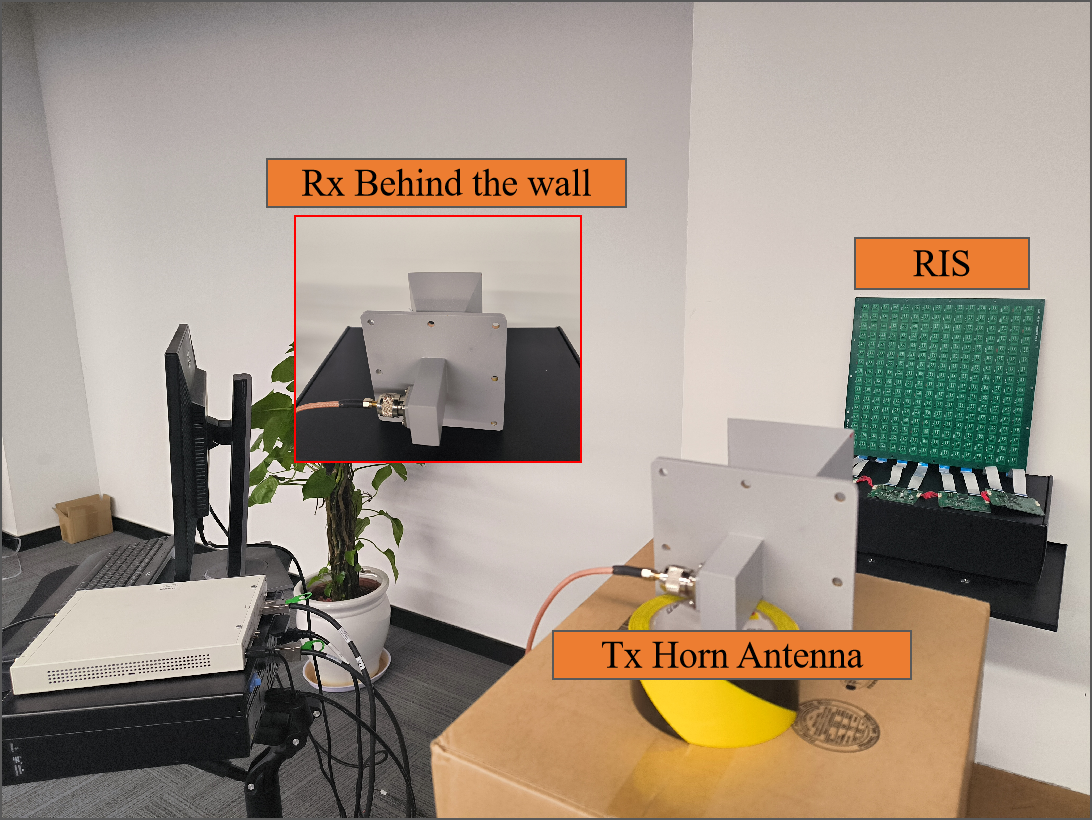}}
  \subfigure[]{
    \label{F:wall_1}
    \includegraphics[width=0.4\columnwidth]{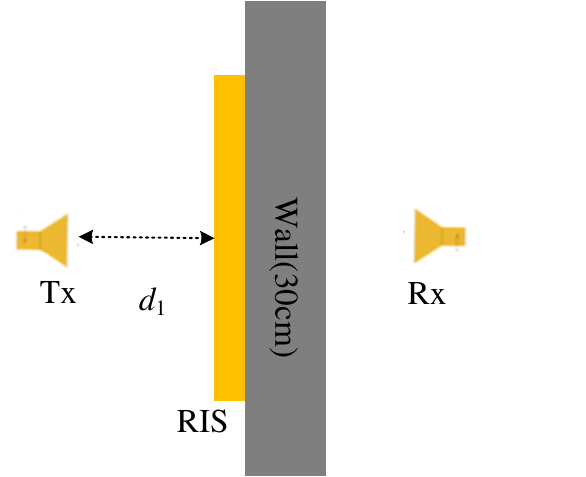}}
\caption{ Transmissive RIS-aided wireless communication prototype, (a) Experimental scene, (b) schematic diagram.}
\label{F:Fig:wall}
\end{figure}
\begin{table}[!t]
\caption{Receiving Power with or without RIS under different $d_1$\label{tab:table2}}
\centering
\begin{tabular}{|c|c|c|c|c|}
\hline
$d_1$ & 0.5 m & 0.8 m & 1.65 m \\
\hline
Received power(without RIS) & -72 dBm & -72 dBm & -67 dBm \\
\hline
Received power(with RIS) & -64 dBm & -65 dBm & -61 dBm \\
\hline
Gain & 8 dBm & 7 dBm & 6 dBm \\
\hline
\end{tabular}
\end{table}
\section{Conclusion}
This work introduces an innovative 1-bit transmissive RIS through the antisymmetry configuration of the two PIN diodes, nearly uniform transmission magnitudes but inversed phase states in a wide band can be obtained. Experimental findings reveal that the transmissive RIS prototype with 16$\times$16 unit cells, achieves a $180^{\circ}$ phase shift and scanning beam at 5.8 GHz. Furthermore, a transmissive RIS-aided communication prototype is established, and its has a good performance in penetrating obstacles.
\section*{Acknowledgment}
This work was supported in part by the Nation Natural Science Foundation of China under Grant No.12141107, and in part by the Interdisciplinary Research Program of HUST, 2023JCYJ012.
\bibliographystyle{IEEEtran}
\bibliography{References}

\end{document}